# REPLAY ATTACK PREVENTION IN KERBEROS AUTHENTICATION PROTOCOL USING TRIPLE PASSWORD

Gagan Dua[1], Nitin Gautam[2], Dharmendar Sharma[3], Ankit Arora[4]

[1]Department of Computer Engineering, National Institute of Technology,
Kurukshetra, India
`dua.gagan@outlook.com`
[2]Department of Computer Engineering, Raj Kumar Goel Institute of Technology,
Ghaziabad, India
`nitin.04it@hotmail.com`
[3]Department of Computer Engineering, Satyug Darshan Technical Campus,
Faridabad, India
`d.sharma000@gmail.com`
[4]Department of Computer Engineering, RIMT-Institute of Engineering & Technology,
Mandi Gobindgarh, India
`cankit087@gmail.com`

*ABSTRACT*

*Replay attack and password attacks are serious issues in the Kerberos authentication protocol. Many ideas have been proposed to prevent these attacks but they increase complexity of the total Kerberos environment. In this paper we present an improved method which prevents replay attacks and password attacks by using Triple password scheme. Three passwords are stored on Authentication Server and Authentication Server sends two passwords to Ticket Granting Server (one for Application Server) by encrypting with the secret key shared between Authentication server and Ticket Granting server. Similarly, Ticket Granting Server sends one password to Application Server by encrypting with the secret key shared between TGS and application server. Meanwhile, Service-Granting-Ticket is transferred to users by encrypting it with the password that TGS just received from AS. It helps to prevent Replay attack.*

## 1. INTRODUCTION

Security in today's world is a major concern. As networks grow, they provide more and more services. Providing these services to the user in a secure way is an issue. Attackers can easily gain information during its transmission across the network and then gain unauthorized access to the servers, to whom they are not able to access. For example, in a distributed environment, nodes or computer are distributed across the network, users want to access services that are stored on servers and servers are distributed. So, in this scenario, servers should be able to authenticate all requests for services. Authentication is a way of ensuring that no one can access the system without providing the way that he has access right. Therefore, instead of each server check request for services, Kerberos provides a central server which does the task of authentication. If an authorized user gains access to the resources, he may either gain access to secret information or may damage resources such as Information stored in the database. Therefore, security is needed at all places in today world from protecting computer resources to the protection of a nation. But security involves implementation of measures to protect attacks. But it does not mean that an attack ill never occur. For example, preventing an outside attacks doesn't' mean that you are secure, attacks may occur from inside of organization. Researchers have proved that many attacks occur from inside of the organization. Therefore, it is necessary to provide security inside of an organization. Authentication protocol is one of the most classical single sign-on protocols. A single sign-on system means that a user can access all services from the application servers after only sign on one time in a multiple application systems. Kerberos V5 is being used at present but there are lots of replay and password attack problems in it [1]. Kerberos V5 was designed to overcome some of the deficiencies of Kerberos V4, but it can't guarantee to avoid replay and password attack. This paper provides triple layer of security. If an attacker successes in gaining access to the ticket-granting-ticket (TGT) and obtaining Ticket-granting-service from Ticket Granting Server (TGS), he will not be able to perform replay attack because authentication server will ask the Ticket-Granting-Service provider (user) about the password.

## 2. RELATED WORK

Many schemes have been proposed to prevent replay attack in Kerberos authentication protocol. Jian [2] proposed an optimized way to prevent password attack and replay attack in single Sign-on system. Multiple databases were added to provide the authentication and authorization in order to prevent replay attack. In this approach, Authentication Server sends Ticket-Granting-Ticket to user as well as to Ticket- Granting-Server (TGS).Similarly; TGS sends Service-Granting-Ticket to both Client and Application server. TGS and Application server, each has their own database. They store these tickets in their database and if attacker replays Ticket-Granting-Ticket (TGT) or Service-Granting-Ticket, they can easily detect whether this is an attack or not

A dynamic double password based sign-on protocol was proposed [3]. That protocol makes use of two passwords that are needed during the user registration and log files concept was used. Log file contained the details when a particular user visited to a server which could be a authentication server, Ticket Granting Server or Application Server. Application server generates log file and forwards to authentication server even after responding the user. Authentication server passes this log file to clients. Similarly, Authentication server also passes its log file. Therefore, a user can make a judgment on security of password through auditing log files and allowed to modifying the password. So, if an attacker has captured a password, client can easily change it by looking and analyzing at the log files.

In [4], a concept is provided to prevent replay attack in Kerberos by using a freshness which makes use of new Symbolic Model Verifier.

Location based Kerberos authentication protocol is described in [5]. In this approach server captures P(Y) code off all the client in the network and it assigns ticket granting ticket to the client by encrypting session key( used for communication between TGS and client) and TGT with the P(Y) code of user. After receiving this message, client accepts its P(Y) code using GPS and decrypts the message. So, if an attacker is able to capture the message, then he will not be able to decrypt the message because P(Y) code length is in several of gigabits. It will result in the failure of the ticket due to time synchronization problems. Here, user physical location is added as an additional message into the Kerberos protocol, which helps to determine physical location of the message provider. Server sends (TGT) to client by encrypting session key with the hash value of user physical location. So, even if an attacker captures a message, he will have to break two phase security to get session ticket and in this process, ticket time may expire.

Capturing user physical location and adding it as a new authentication factor into the Kerberos Protocol method [7] was proposed to prevent replay attack. It used N-BAN logic (modified version of BAN logic [6]) to apply on the modified Kerberos protocol.

Benjamin [8] proposes a method for the inspection of replay attacks on Kerberos authentication protocol in which the protocol was specified by using the Object-Z.

Modified Symbolic Model verifier [9] approach was presented to find problems with respect to the replay attack.

Some basic principles [10] were defined which are necessary to be used while designing the cryptography protocols. Five different strategies are presented. By using these strategies it is possible to design cryptographic protocols which show robustness against different classes of replay attacks.

A new protocol for key distribution was proposed [11] after analysing the security flaws with different protocols that are currently used for the authentication as well as for key distribution. This proposed model is based on using symmetric keys.

This paper is further organized into the following sections: Section III provides an overview about the Kerberos authentication protocol. The Section IV provides some limitations about Kerberos authentication protocol. In section V, we proposed our new method to prevent replay attack in Kerberos environment which makes use of three passwords from the users and look how this architecture provides protection against password and replay attack.

## 3. OVERVIEW OF KERBEROS PROTOCOL

Kerberos is an authentication protocol which is used to authenticate users in the distributed environment. Using Kerberos authentication protocol, a client can authenticate itself to multiple servers using its password which is also known as the long term secret key. Client receives Ticket-Granting-Ticket (TGT) from the Authentication Server (AS) and this ticket can be used for multiple services that a client needed. Therefore, client stores this Ticket-Granting-Ticket in its database. Then, it requests for Service-Granting-Ticket and stores it in its database[12].An advantage of storing Service-Granting-Ticket in database is that client will not have to re-enter password every time when he has to access the application server such as email server. A Kerberos environment consists of Key Distribution Center (KDC), a number of clients and Application Servers. Key Distribution Center (KDC) consists of authentication server (AS) and Ticket Granting Server (TGS). An AS issues Ticket-Granting-Tickets to the user after the verification and TGS issues Service-Granting-Tickets to the user. If a client wishes to authenticate herself to application server, then Kerberos will perform this task in three phases as:

1. Whenever a user logs on to a workstation, the client process running in the workstation sends a message to the Authentication Server. Authentication Server checks in the database whether username and password are correct. If all is correct, then Authentication Server (AS) sends a Ticket-Granting-Ticket to the user and also a session key so that user can communicate with the server. Same copy of session key is also included in the ticket that AU issues to the client. Ticket-Granting-Ticket and session key are encrypted using a key generated from the user password. Because message sent by the AU to client is encrypted using key generated from user password, only authorized user reads the

      message. Client decrypts the message and gets Ticket-Granting-Ticket that is to be used for communication with the Ticket Granting Server (TGS).

2. Now, Client presents this Ticket-Granting-Ticket (TGT) to the Ticket Granting Server along with an authenticator. Authenticator simply includes the ID of the user on client and also a timestamp. As compared to a ticket, an authenticator has very short life time. Since, Ticket-Granting-Ticket (TGT) is encrypted with a secret key that is shared between Authentication Server and Ticket Granting Server (TGS). Only TGS can decrypt the message. After decrypting the message it gets a session key which is used to decrypt authenticator received from the user. Now, TGS checks the sender details by comparing TGT with the details in the authenticator and incoming packet network address. If all details are verified, TGS issues a Service-Granting-Ticket to the client and it is encrypted using the secret key shared between TGS and application server.TGS also generates a session key to be shared between client and application server for secure communication. TGS sends this session key to Server by encrypting it in the ticket and also to the client by encrypting the message with the session key that was shared between client and TGS.

3. In the third step, client presents this ticket to the server along with an authenticator. Client encrypts the message by using session key that was sent by the TGS. Server then uses that session key to encrypt the message from the client. If all credentials of the user are correct, then application server will issue a response to the client in case if a mutual authentication is required [13].

Figure 1 shows basic Kerberos architecture. Here, directions of arrows show how data will flow in the Kerberos environment. Meaning of each number is:

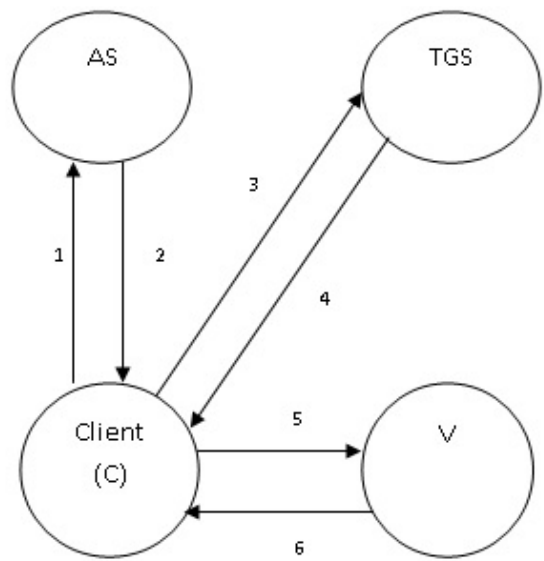

Figure1. Basic Kerberos Architecture

1. Request for Ticket-Granting-Ticket

2. Ticket-Granting-Ticket + Session Key

3. Request for Service Granting Ticket

4. Service-Granting-Ticket + Session Key

5. Request Service

6. Provide Service authentication

Meaning of the terms used in the figure is described in the table1 which is given below:

Table1: Meaning of the terms used in Figure1

| Term | Meaning |
|------|---------|
| C | Client |
| AS | Authentication Server |
| TGS | Ticket Granting Server |
| V | Application Server |

These six steps form basic Kerberos architecture. The main problem with Kerberos Authentication protocol is of replay attack. Replay attack is an attack in which attacker captures messages transmitting through the channel, modifies it and replay back on the transmitting channel [14]. So, it is necessary to prevent the replay attack especially when two parties need secure communication over the internet.

## 4. KERBEROS LIMITATIONS

Although Kerberos is being widely used, but it has a number of limitations which are described below:

1. Kerberos authentication protocol requires continuous availability of the key distribution server (KDC). If KDC fails, then the entire Kerberos environment will fail.
2. User must choose strong password. A selection of weak password could result in the password or replay attack. An attacker can easily break weak password and gain access to the services stores on the server.
3. System clocks of the clients, Authentication server, key distribution Center and client should match; otherwise timely attack can be easily performed on the Kerberos protocol.

Kerberos makes use of KDC (Key Distribution Center), KDC uses Kerberos database where details of all the users are stored. An attacker can perform an attack on this database and may gain access to the database [15, 16(P 87)].

## 5. PROPOSED MODIFICATIONS TO THE KERBEROS PROTOCOL

The main problem with the Kerberos Authentication Protocol is that of replay and password attack. Problem arises when Authentication Server (AS) sends Ticket-Granting-Ticket (TGT) to the client process running in the user. Kerberos V5 even can't avoid the replay attack. An attacker can capture all the messages transmitting from the Authentication Server (AS) to the user and apply all possible combination on the messages that he has captured. After applying all the possible combination of the captured messages, an attacker presents TGT to the Ticket-Granting-Server (TGS). TGS checks that this is a valid authenticator, so it passes Service-Granting-Ticket to the attacker and attacker may gain unauthorized access to the services stored on the Application Server (which is V in our Figure1).

The proposed Kerberos authentication protocol is shown in Figure 2. In this modified Kerberos protocol, Ticket-Grating-Server (TGS) presents Session Key to the server by encrypting them encrypting them with the hash of user password.AS passes two passwords to the TGS and TGS further passes one password to the Application Server V.

Our Proposed architecture works as follows:

> First, a user logs on his workstation and enters the password. If there is the new user then he will have to enter three passwords to complete his registration. After entering the passwords, the client process running in the workstation sends a message to the Authentication Server. This message consists of the Ticket request that client process requires for communication with the Application server (V).Authentication Server checks in the database whether user credentials are correct. If these are correct, then Authentication Server sends Ticket-Granting-Ticket to the client process. Authentication server stores three passwords in its database.
>
> After responding to the user, Authentication Server sends two user passwords to the TGS.
>
> Client presents Ticket-Granting-Ticket to the TGS along with an authenticator. Here, it may possible that this ticket and authenticator came from the attacker. TGS decrypts the TGT and gets session key which will be used for further communication with the client and also it gets user passwords by decrypting message received from AS with its secret key.
>
> TGS produces Service-Granting-Ticket and session key to be shared between client and Application Server. It encrypts the session key with the user password that it received from the AS. This is very important step because only the valid user will be able to decrypt the message. If that message to TGS came from the attacker, he will not able to continue furthest communication. TGS also sends user password to the server by encrypting it with the secret key that is shared between the TGS and Application Server.
>
> Client finally presents Service-Granting-Ticket to the Application Server V; it will also send an authenticator by encrypting it with the session key that it just received from the TGS. Server will receive the message and cheek the credentials of the client with the help of authenticator that it received from the client. To enhance the security, server will ask from the user about the password. If this is the client who just sent message to server, he will immediately tell his password to server by encrypting it with the session key shared between client and the Application Server. Using session key here is more advantageous because whenever user will login to his workstation, it will be changed every time.
> Server will set a timer at its side and it is used for the detection of any attack. If password from the client does not arrive within the specified time, then it will send a message to TGS to tell him that the ticket you just issued had been gained by the attacker.

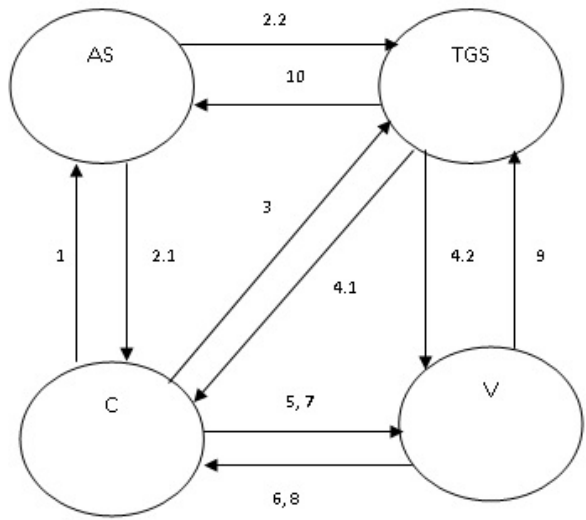

Figure2. Proposed Kerberos Architecture

Table2 below shows various Kerberos messages that are used in this proposed method. These messages are divided into different categories:

1. Messages for obtaining Ticket-Granting-Ticket
2. Messages for obtaining Service-Granting-Ticket
3. Client-Server Messages to obtain Service

**Let us now describe the step-by-step working of our proposed algorithm**:

(1) During step1 of the algorithm client sends a message to the AS. When client logs on first time on the workstation, then he will have to enter three passwords. These passwords will be stored in a database that will be maintained at the AS. Thereafter client will have to enter only one password. Every time, a user logs on to his workstation, its credentials are checked against the information maintained at AS. If user credentials are correct, then a process running in the user workstation will send a message to the AS. This message consists of the identification of the client and the Server with which the client wants to communicate.

(2.1) At this step AS sends TGT to the client. It also sends the session key ($K_{c,tgs}$) that will be used for communication between the AS and TGS and this session key ($K_{c,tgs}$) is encrypted with the user password ($K_{c1}$) so that only the authorized user can get the session key.

(2.2) At this step of the algorithm, AS presents two of the passwords ($K_{c2}$, $k_{c3}$) from its database to the Application server V. As you can see, these passwords are encrypted with the secret key ($K_{tgs}$) that is known only to the AS and TGS.

Table2. Proposed Kerberos Messages exchange

**(a) Steps to obtaining Ticket Granting Ticket:**

(1) C → AS: $ID_c \| ID_{tgs} \| N_1 \| times$

(2.1) AS → C: $ID_c \| ticket_{tgs} \| E(K_{c1}, [K_{c,tgs} \| ID_{tgs} \| N_1 \| times])$
where $ticket_{tgs} = E(k_{tgs}, [ID_c \| AD_c \| times \| K_{c,tgs}])$

(2.2) AS → TGS: $E(K_{tgs}[K_{c2}, K_{c3}])$

**(b) Steps for obtaining Service Granting Ticket:**

(3) C → TGS: $ticket_{tgs} \| ID_v \| authenticator_c$
where $authenticator_c = E(K_{c,tgs}[ID_c \| AD_c \| TS_2])$

(4.1) TGS → C: $ID_c \| ticket_v \| E(K_{c2}, [N_2 \| ID_v \| K_{c,v} \| ID_v])$
where $ticket_v = E(K_v, [K_{c,v} \| ID_c \| AD_c \| times \| T_3])$

(4.2) TGS → V: $E([K_v, K_{c3}])$

**(c) Client-Server authentication messages for obtaining service:**

(5) C → V: $Ticket_v \| authenticator_c$

(6) V → C: Setup timer & $E(K_{c,v}, [ID_c \| K_{c3}])$

(7) C → V: $E[K_{c,v}[K_{c3} \| T_5]]$
(8) V → C: $E(K_{c,v}[T_5 + 1])$

(9) V → TGS: if response does not come, then tell TGS about the address from where attack occurred.
(10) TGS → AS: Forward message to AS to inform about the attack.

(3) Because authorized user knows the password, so it gets the session key by decrypting the message, in this step, client generates an authenticator that consists of network address from where message came as well as the timestamp value and this authenticator is encrypted with the session key ($k_{c,tgs}$). Client presents this authenticator with the TGT that it received from AS in step 2.1

(4.1) After receiving the ticket in step3, first TGS decrypts it using its private key ($K_{tgs}$). After decryption, it gets the session key ($K_{c,tgs}$). Using this session key, TGS decrypts the authenticator. Now, TGS produces the Service-Granting-Ticket which is encrypted with the secret key ($K_v$) shared between TGS and V. TGS also produces the session key ($k_{c,v}$) that is sent to the user by encrypting with the user password ($K_{c2}$) that TGS received from AS in step 2.2. This is very important step because if an attacker gains success in capturing the TGT, h will not be able to replay further messages to the Application Server (V) because only the authorized user knows the password ($K_{c2}$). This step prevents the replay attack.

(4.2) At this step, TGS sends $k_{c3}$ (user password) to V by encrypting it with the secret key $K_v$

>that is shared between the TGS and Application server V. After receiving this message, Application Server (V) decrypts it using the $K_v$ that is private to V. Sending of this password will help in detection of attacks such as replay of messages and gaining access to the user passwords.

(5) Client presents Service-Granting Ticket as well as authenticator to the Server V. At this step, it may possible that Ticket came from an authorized person or from attacker.

(6) To verify identity of the sender, server asks about the password ($K_{c3}$) from the user who sent the message. As Server V sends a message to the user asking about the password, it sets up a timer. We can see this message will be encrypted with the session key $K_{c,v}$.

(7) If sender is the authorized party, it will immediately tell password to the server and this password will be sent to the server by encrypting it with the session key ($K_{c,v}$).Client will also send timestamp value indicating the time at which this response was generated.

(8) If sender is confirmed that message came from legitimate client, it will provide response to the client by incrementing timestamp by 1.this is necessary in case a client required a mutual authentication. Mutual authentication is requires in which server have also to prove his/her identity to the user. As we can see that Server V provides incremented value of the timestamp {$T_5+1$} to the user because client sent timestamp value $T_5$ in the step5 to the Server V.

(9) If server does not receives the response before its timer gets off, it will be conformed that the Service-Granting-Ticket came from an attacker. So, Application Server (V) sends an alert message to the AS informing it about the address from where attack occurred.

(10) Now, TGS further informs to AS about the attack and now, it is the responsibility of the Authentication Server to inform about the attack. If user password has been compromised, then AS will inform about this to the user.

So, this architecture can detect whether user password has been compromised. Replay of messages can also be detected by this architecture.

### 5.1 How this Architecture prevents from the attacks

Suppose an attacker has applied all the possible combinations and made a guess of the session key $K_{c,tgs}$. Let us look how messages can be replayed by the attacker.

**Attack1. When attacker captures the session key: $K_{c,tgs}$**

Suppose an attacker has captured the session key ($K_{c,tgs}$), then he can easily replay messages to the TGS by hacking the client location. Now, question is that how TGS prevents from this attacker. Instead of encrypting the session key $K_{c,v}$ (that is to be shared between the client and Application Server V) with the $K_{c,tgs}$ (that has captured or guessed by the attacker), TGS sends it by encrypting it with the user password ($K_{c2}$) that is known only to the user. So, if this ticket was generated from the attacker, he will not be able to use the service from the Application Server (or will not be able to replay messages to the Application Server V) because attacker does not know the user password ($K_{c2}$).

**Attack2. When attacker have guessed user password: $K_{c2}$**

Second problem is that how this architecture prevents from the attack that can occur when an attacker has made guessed of the user password $K_{c2}$. (note that if an attacker has made an guess of the user password $K_{c2}$, then he can easily get access to the session key $K_{c,v}$ that is to be shared between the client and application server).

If attacker has access to $K_{c2}$, then he can easily perform replay attack on the Application Server. Our architecture can easily prevent this attack. Now, it is the responsibility of the Server V to detect the presence of replay attacker.

Suppose, in step 5 it was the attacker who replayed an message to the Application server, Now, server will ask about the $K_{c3}$. An attacker does not know $K_{c3}$. Server will use timer to get the response within the specified time.

So, as we can see our architecture provides triple layer of security. First level of security is provided by $K_{c,tgs}$. This session key will be renewed every time user requires access to a new service that is stored on a different Application server. Second level of security is provided by the user password $K_{c2}$ stored in the database at TGS. Third level of security is provided by the $k_{c3}$ which has already been forwarded to the Application server during step 4.2 of the algorithm.

### 5.2 Advantages of the Proposed Architecture

(i)   Provides protection against replay attacks
(ii)  It provides triple layer of protection.
(iii) Password attack protection

### 5.3 Limitation of the Proposed Architecture

(i) When a new user logs on the workstation, then he will have to enter three passwords and these passwords will be stored at the TGS.

### 5.4 Comparison with the Figure1

Kerberos architecture shown in the figure1 is the basic Kerberos architecture. Replay and password attacks can easily be performed in the architecture shown in figure1. Architecture shown in the figure1 also provides only single level of protection by using session keys that are renewed every time user logs on to the workstation and requests for service.

But our architecture provides protection against replay as well as against password attack. Our architecture provides triple level of security as compared with the single level of security. If a user password has been compromised then architecture shown in figure1 does not address this issue. Bur using our architecture, user can be informed about the attack and user can take necessary steps for making change in the password.

## 6. CONCLUSION

Security is necessary in all aspects of fields. Kerberos provides third party authentication but many replay attacks occurred on Kerberos. The approach used in this paper attempts to prevent replay attack by using three passwords, a new user must enter these passwords that will be stored on the Authentication Server. If an attacker gains access to TGT, then he can easily replay them to the TGS, but not to the Application Server (V). The reason for this is that attacker does not know the password to get session key used for communication with the Server V. So, we have to prevent attacks from taking unauthorized take control from system even if he has gain access to session key and the ticket. The approach used in our proposed architecture provides protection against replay and password attack.

International Journal of Computer Networks & Communications (IJCNC) Vol.5, No.2, March 2013## Authors

**Gagan Dua** is working as Assistant Professor in the Department of Computer Engineering at National Institute of Technology, Kurukshetra. He completed the Masters in Computer Science & Engineering from Jaypee University of Information Technology, Solan in 2012. He received the Bachelor's Degree in Computer Science & Engineering in 2009. His research interest includes Network Security, Parallel and Distributed Computing and Microprocessor.

**Nitin Gautam** is working as Assistant Professor in the Department of Computer Engineering at Raj Kumar Goel Institute of Technology, Ghaziabad. He completed the Masters in Computer Science & Engineering from Jaypee University of Information Technology, Solan in 2012. He received the Bachelor's Degree in Information Technology in 2009. His research interest includes Security and Scheduling in Wireless Sensor networks.

**Dharmendar Sharma** is working as Assistant Professor in the Department of Computer Engineering at Satyug Darshan Technical Campus, Faridabad, He completed the Masters in Computer Science & Engineering from Jaypee University of Information Technology, Solan in 2012. He received the Bachelor's Degree in Information Technology in 2009. His research interest includes Data Mining and Data Warehousing.

**Ankit Arora** is an M.Tech. Student in the Department of Computer Engineering at RIMT Institute of Engineering & Technology, Mandi Gobindgarh, He received the Bachelor's Degree in Computer Science & Engineering in 2009. His research interest includes Networks and parallel Algorithms.70